\documentclass[conference]{IEEEtran}
\IEEEoverridecommandlockouts
\usepackage{cite}
\usepackage{amsmath,amssymb,amsfonts}
\usepackage{algorithmic}
\usepackage{graphicx}
\usepackage{textcomp}
\usepackage{svg}
\usepackage{xcolor}
\usepackage{enumitem}
\usepackage{parskip}
\usepackage{multicol}
\usepackage{multirow}
\def\BibTeX{{\rm B\kern-.05em{\sc i\kern-.025em b}\kern-.08em
    T\kern-.1667em\lower.7ex\hbox{E}\kern-.125emX}}

\begin{document}

\title{FPGA Divide-and-Conquer Placement \\using Deep Reinforcement Learning
}

\author{
	Shang Wang$^{1}$, Deepak Ranganatha Sastry Mamillapalli$^{1}$, \\ Tianpei Yang$^{1}$,
	and Matthew E. Taylor$^{1,2}$\\	
	$^1$University of Alberta, Canada\\
         $^2$Alberta Machine Intelligence Institute (Amii), Canada\\
        \texttt{\{shang8,mamillap,tianpei.yang,matthew.e.taylor\}@ualberta.ca}
}

\maketitle

\begin{abstract}
This paper introduces the problem of learning to place logic blocks in Field-Programmable Gate Arrays (FPGAs) and a learning-based method. In contrast to previous search-based placement algorithms, we instead employ Reinforcement Learning (RL) with the goal of minimizing wirelength. 
In addition to our preliminary learning results, we also evaluated a novel decomposition to address the nature of large search space when placing many blocks on a chipboard.
Empirical experiments evaluate the effectiveness of the learning and decomposition paradigms on FPGA placement tasks. 

\end{abstract}


\section{Introduction}
\begin{figure}[thp]
\centerline{\includegraphics[width=0.8\linewidth]{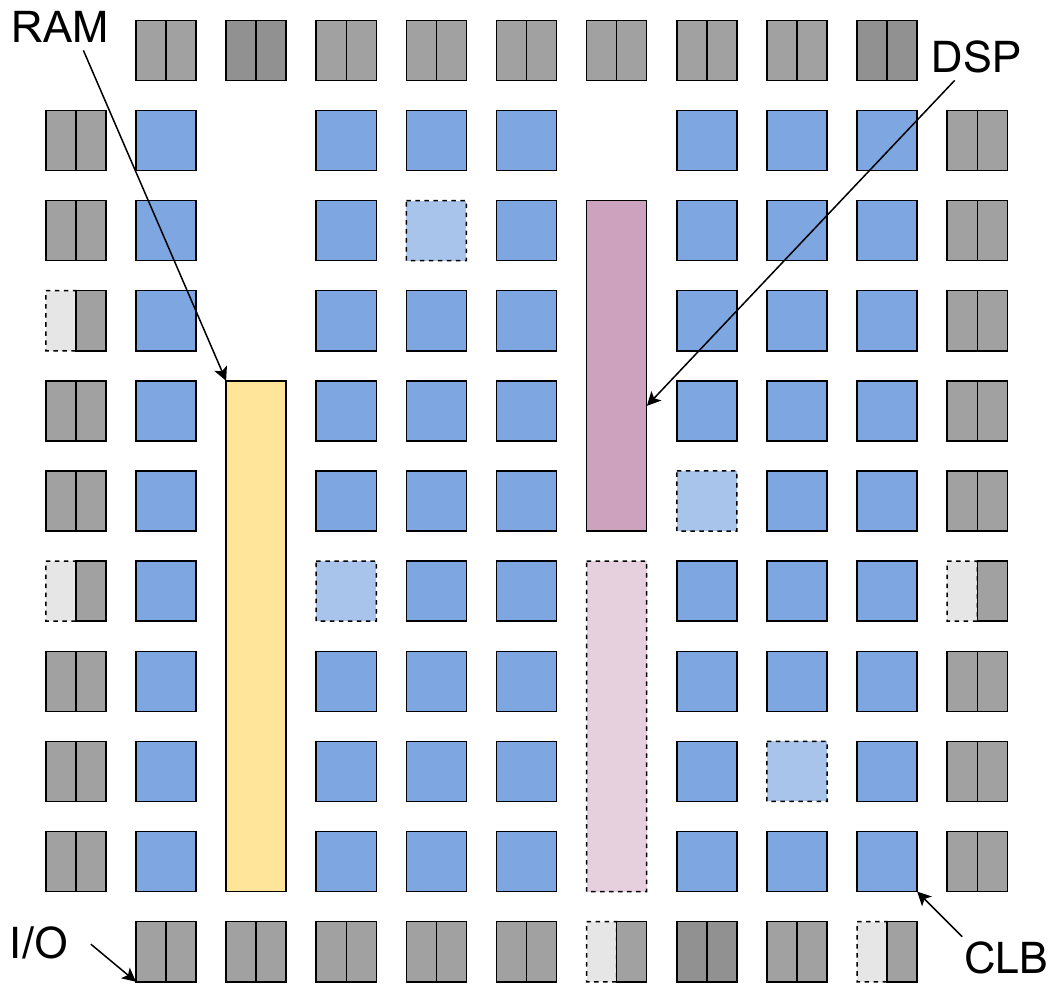}}
\caption{The FPGA board is $11 \times 11$ units in size and incorporates DSP (Digital Signal Processor), CLB (Configurable Logic Block), I/O, and RAM (Random Access Memory) blocks, as well as I/O locations of capacity 2.}
\label{board}
\end{figure}

The relentless advancement in very-large-scale integration, characterized by the increasing scale and complexity of chips, intensifies the challenge in Electronic Design Automation (EDA). Optimizing the arrangement of thousands of circuit modules within strict design constraints (e.g., the placement problem) becomes a particularlly challenging task. 
Low-quality placement results can lead to challenges for successive processes, e.g., routing, and can strain the limited capacity of FPGAs prefabricated routing resources, critically affecting the design's final performance. However, FPGA placement is a more constrained problem than ASIC placement because block types and wiring patterns are prefabricated (see Fig.~\ref{board}).

Simulated annealing (SA) \cite{kirkpatrick1983optimization} has traditionally been the backbone algorithm in FPGA placement methods, respecting placement constraints and minimizing routing delays. Combining analytical placement (AP) and SA \cite{gort2012analytical, chen2017ripplefpga} is currently the dominant technique used in placement optimization. For example, the commercial Intel Quartus placer \cite{IntelQuartus} uses AP to determine the initial placement solution and SA to fine tune the placement. RLplace \cite{elgammal2021rlplace}, the current state-of-the-art approach, enhances SA by allowing a reinforcement learning (RL) agent to choose from multiple types of directed moves. 
However, RLplace still relies on simulated annealing, which can be slow to converge, particularly for complex FPGAs. Additionally, the bandit formulation assumes that rewards are based solely on intrinsic properties, lacking contextual information.

AI has gained traction to accelerate chip design \cite{mirhoseini2020chip,cheng2021joint,lai2022maskplace,chipformer}. 
RL addresses placement as a sequential decision-making challenge, placing each circuit module at a time, producing chip layouts that match or surpass those designed by humans or traditional algorithms, particularly in metrics like wirelength and congestion. However, these advances have mainly focused on ASIC, while FPGA placement represents a more constrained problem. Moreover, in both FPGA and ASIC, RL agents face significant challenges, including extremely sparse rewards and vast search spaces. An RL agent may find it difficult to identify and reinforce beneficial behaviors when feedback is infrequent and the required exploration is enormous \cite{abs-2109-06668}.\footnote{As pointed out elsewhere \cite{mirhoseini2020chip}, the state space of the placement task is larger than $10^{2500}$ with only 1000 blocks, while Go (an extremely challenging RL benchmark) has an estimated state space of $10^{170}$.}

This proposes a RL approach to address the FPGA placement challenge. Distinguishing our approach from RLplace \cite{elgammal2021rlplace}, we model the FPGA placement as a Markov decision process (MDP), rather than as a multi-armed bandit. Our method does not rely on conventional heuristic search algorithms, potentially enhancing the efficacy and adaptability of our approach in the FPGA setting. This paper has two main contributions: 1) 
we apply RL to FPGA placement by introducing a novel state definition and model architecture. 2) To tackle the immense search space inherent in placement tasks, we propose a new training paradigm that decomposes the full placement problem into smaller and more manageable subtasks, improving the agent's learning efficiency. To the best of our knowledge, we are the first to introduce a (deep) RL agent for FPGA placement and the first to propose a divide-and-conquer approach to optimize chip placements, with preliminary results that suggest the approach's feasibility. We hope that this paper will attract more attention to the use of RL in FPGAs by electronic design automation engineers.

\section{Preliminaries}

\subsection{FPGA Placement} The chip placement problem maps netlist components, a list detailing a circuit's components and connections, onto the chipboard. The number of components to be placed in FPGA problems can range from hundreds (for simple circuits) to tens of thousands (for complex, high-density designs), illustrating the scale and variability of the placement challenge. Circuit modules positioning, like CLBs and I/Os, must respect two constraints. 1) Type constraints state that blocks can only be placed in certian (prefabricated) locations. 2) Capacity constraints state that blocks can only be placed within capacity limits per position. For example, in Fig.~\ref{board}, the next CLB block can only be placed in a dotted blue square, since only the blue square positions accept CLB blocks and have additional capacity. The most common goal in FPGA placement is to minimize wirelength and critical path delay. We focus on wirelength as our primary metric to simplifying our assessment. Wirelength refers to the total distance that interconnects cover on a circuit board, summarizing the cumulative path lengths of wires needed to connect components. In our case, the wirelength is generated by VTR (Verilog-to-Routing) \cite{vtr2022benchmarks} after routing. VTR is an open-source CAD tool that provides a complete suite of tools for FPGA design, including synthesis, mapping, placement, routing, and timing analysis. 

\subsection{Reinforcement Learning} RL is a machine-learning paradigm where an agent learns to make decisions by performing actions in an environment, learning to maximize (discounted) cumulative rewards. This learning process is often modeled as a Markov decision process (MDP), which describes a problem in terms of states, actions, rewards, and state transitions. States represent the possible situations the agent can encounter. Actions are available decisions that an agent can execute. Rewards are numerical values that the agent receives as feedback after taking actions. State transitions define how actions change the state.

The actor-critic framework is one common method to learn how to act, integrating the actor for making decisionss and the critic to evaluates those decisions. The actor learns a policy, $\pi(a|s; \theta)$, where $a$ is the action chosen in state $s$ and $\theta$ are the weights of the policy neural network (actor). 
The critic evaluates the quality of the actions by estimating a value function, $\hat{V}(s; \chi)$, where $s$ represents the state and $\chi$ indicating the weights of the value neural network (critic). Learning adjusts the weights of both the actor and the critic to improve the policy and the value function estimations, respectively. The actor updates its policy using gradients estimated from the critic’s evaluation, and the critic updates its value approximation by minimizing the temporal difference error. 

\begin{figure*}[thp]
\centering
\includegraphics[width=0.9\linewidth]{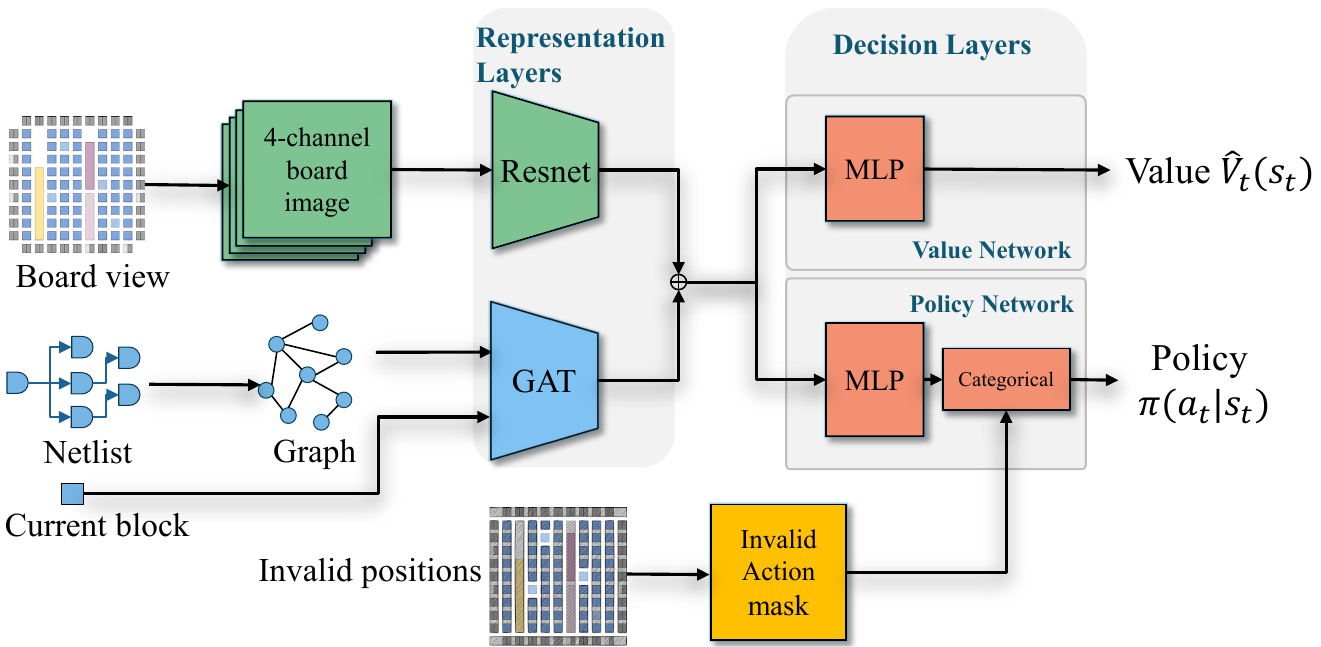}
\caption{Overview of our model structure, which contains two main parts: representation layers and decision layers, including a policy network, and a value network. The representation layers take board observations, the netlist graph, and the current block index as input, while the decision layers output a probability distribution over available placement locations (the policy $\pi(a_t|s_t)$) and an estimate of the expected reward for the current placement (the state value $\hat{V}_t$).}
\label{arch}
\end{figure*}
\section{Our Approach}


\subsection{Problem Formulation} 
We have chosen to formulate the FPGA placement problem as an MDP. Our MDP consists of four components: \textbf{1) States} consist of the netlist graph, the current block to be placed, and the placement status of the current board. \textbf{2) Actions} are defined as locations where a given block can be placed without violating hard constraints (e.g., capacity or block type limits). \textbf{3) Rewards} are formulated so that intermediate steps receive a reward of 0 until the full placement is completed, at which point the final reward is based on the true wirelength generated by VTR \cite{murray2020vtr} after routing. \textbf{4) State transitions} are determined by the new board layout after placing a block.

\subsection{State Construction} 
We define the state as an image composed of four channels based on the current state of the board and information from the netlist graph about the block to be placed. The four channels are matrices of the same size as the chipboard:
\begin{itemize}
\item The capacity channel indicates the remaining capacity of each grid cell.
\item The input channel indicates the number of times the placed block serves as a source/input in all nets. 
\item The output channel indicates the number of times the placed block serves as a sink/output in all nets.
\item The wire-mask channel \cite{lai2022maskplace} represents how the estimated wirelength increases if a block is placed in a position. 
\end{itemize}
In addition to the board observations, the netlist graph is an undirected graph that encodes the connections between various blocks and the nets that link them. To extract from this graph, we generate a vector representation for each node by concatenating node-specific features, encompassing the node's type, index, and x and y coordinates. 

\subsection{Learning Algorithm} 
We employ the actor-critic algorithm Proximal Policy Optimization (PPO) \cite{schulman2017proximal} to train a neural network
(see Fig.~\ref{arch}). The board layout information is processed through ResNet \cite{he2016deep}, while the netlist graph is handled by the Graph Attention Network (GAT) \cite{velickovic2017graph} to embed node representations. The board observation embedding and the current block embedding vectors are then concatenated to form the state embedding. Notably, we need to filter out invalid actions from a probability action matrix produced by policy networks to adhere to FPGA placement constraints. Others \cite{DBLP:conf/flairs/HuangO22} have examined the effectiveness of invalid action masking, demonstrating that it enables agents to learn more efficiently within large discrete action spaces. Therefore, before sampling actions, we remove invalid actions
by applying invalid action masks.
Over time, PPO updates the policy $\pi_\theta(a_t|s_t)$ by using an estimate of the advantage $\hat{A}_t = G_t - \hat{V}_t$ on time step $t$. $G_t = \sum^{T-t-1}_{k=0}\gamma^k r_{t+k+1}$ denotes the cumulative discounted reward (to be maximized) and $\hat{V}_t$ is the value approximation produced by the value network. 


%
%
%

\section{Experiments}

\begin{figure*}[th]
\centering
    \includegraphics[width=\linewidth]{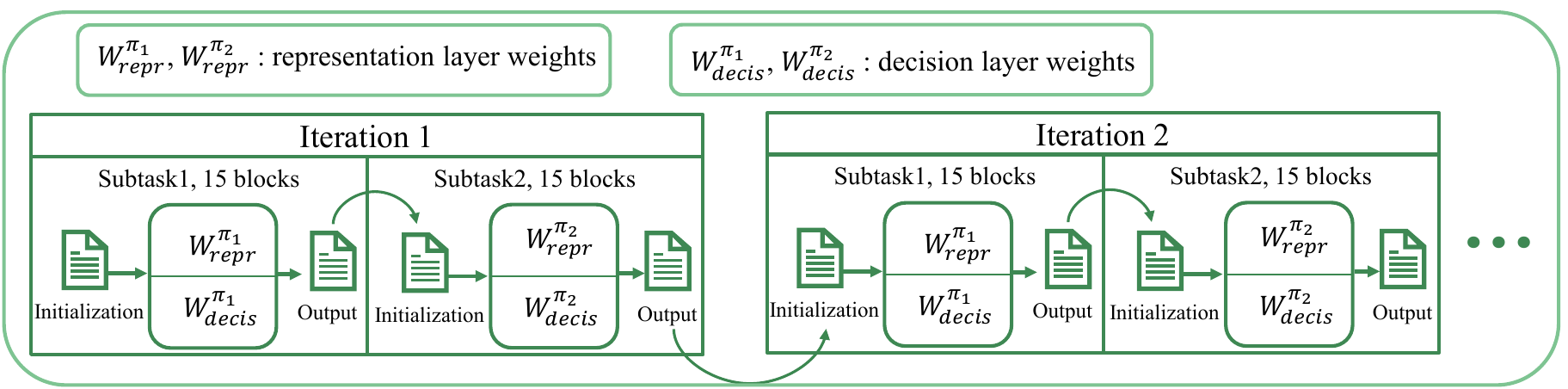}
\caption{The 30-block decomposition training paradigm.}
\label{decomposition}
\end{figure*}

In this section, we evaluate our RL agent under the base and the decomposition setting. We employ the \texttt{tseng.net} netlist and \texttt{EArch.xml} architecture files from MCNC20 benchmarks \cite{Yang1991LogicSA} included in VTR \cite{vtr2022benchmarks}, commonly  used in FPGA design. 
This setup consists of 56 CLBs and 174 I/O blocks. In our initial experiments, we found that the agent struggled to place all the blocks, yielding results no better than random placement. As a result, we decided to simplify the experiments (see Table~\ref{tab1}). Assuming VTR-derived placements are optimal, we start by placing a subset of blocks based on a VTR optima, but allow the agent to learn to place 5 CLBs.
This allows us to assess the model's learning capabilities by examining the performance gap between the VTR-optima and the RL placement results. A narrower gap indicates the RL agent's proficiency in placing the subset of blocks. This process incrementally expands the subset size until it can effectively place all blocks. We compare our RL agent's placements to VTR's, as it is the current state-of-the-art baseline in the placement and routing of FPGAs.

\begin{table}[th]
\begin{minipage}{\columnwidth}
    \centering
    \begin{center}
\caption{One benchmark has 56 CLBs and 174 I/Os blocks \cite{vtr2022benchmarks}. We assume VTR-derived placements are optimal, which is 6489. We remove a subset of blocks from this optimal configuration and allow an RL agent place them.}
\begin{tabular}{|p{2.0cm}|p{1.2cm}|c|}
\hline
\textbf{\# blocks placed by RL agent} & \textbf{avg wirelength}  & \textbf{possible configs}\\
\hline
5& 6398$\pm$71 & $10^5$\\
\hline
15& 6531$\pm$132  & $10^{17}$\\
\hline
30& 6988$\pm$98 & $10^{39}$\\
\hline
45& 7209$\pm$168 & $10^{64}$\\
\hline
56& 7591$\pm$109 & $10^{83}$\\
\hline
\end{tabular}
\label{tab1}
\end{center}
\end{minipage}\hfill 
\end{table}

\subsection{15 Block Decomposition} 
We found the RL agent had difficulty with the very large state space and sparse environment rewards. Table~\ref{tab1} demonstrates that the RL agent can approach the VTR optima in small tasks. However, with an increasing number of blocks, the search space expands exponentially, making the RL agent more prone to converging to suboptimal solutions. To address this issue, we propose a divide-and-conquer approach. As evident from Table~\ref{tab1}, when placing less than 15 blocks, the agent can optimize placement to close or even surpass VTR optima. Therefore, we decompose the full placement problem with a vast search space into subtasks with smaller search spaces. 

Using Fig.~\ref{decomposition} as an of placing 30 blocks, the task will be decomposed into two subtasks with 15 blocks each. One training iteration includes training procedures of the two subtasks. Once one subtask is trained, its placement results will be fixed as the initialization placement for the other subtask to place the other 15 blocks. For each subtask’s model, we propose four settings:

\begin{enumerate}
\item Setting 1: multi-policies, reusing all weights\\
Each subtask has separate model weights to initialize the policy and all model weights of each subtask are reused in the next iteration.
\item Setting 2: multi-policies, not reusing decisions layers weights\\
Each subtask has separate model weights to initialize the policy. Only the representation layer weights of the models are reused in the next iteration. 
\item Setting 3: single policy, reusing all weights\\
All subtasks share the same model weights and all model weights are reused in the next iteration.
\item Setting 4: single policy, not reusing decisions layers weights\\
All subtasks share the same model weights and only the representation layer weights of the model are reused in the next iteration. 
\end{enumerate}

\begin{figure*}[ht]
\centering
    \includegraphics[width=0.65\columnwidth]{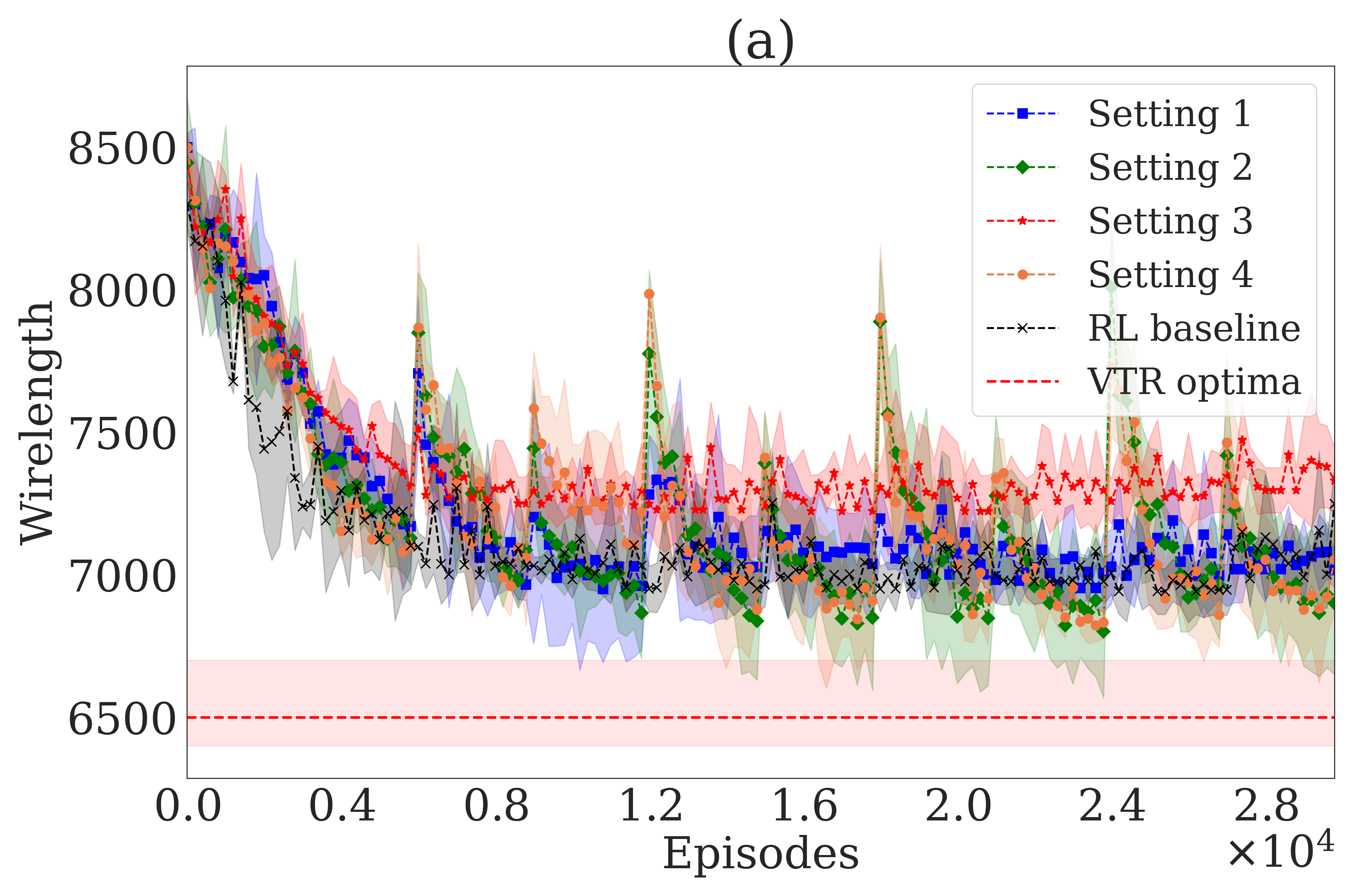}
    \includegraphics[width=0.65\columnwidth]{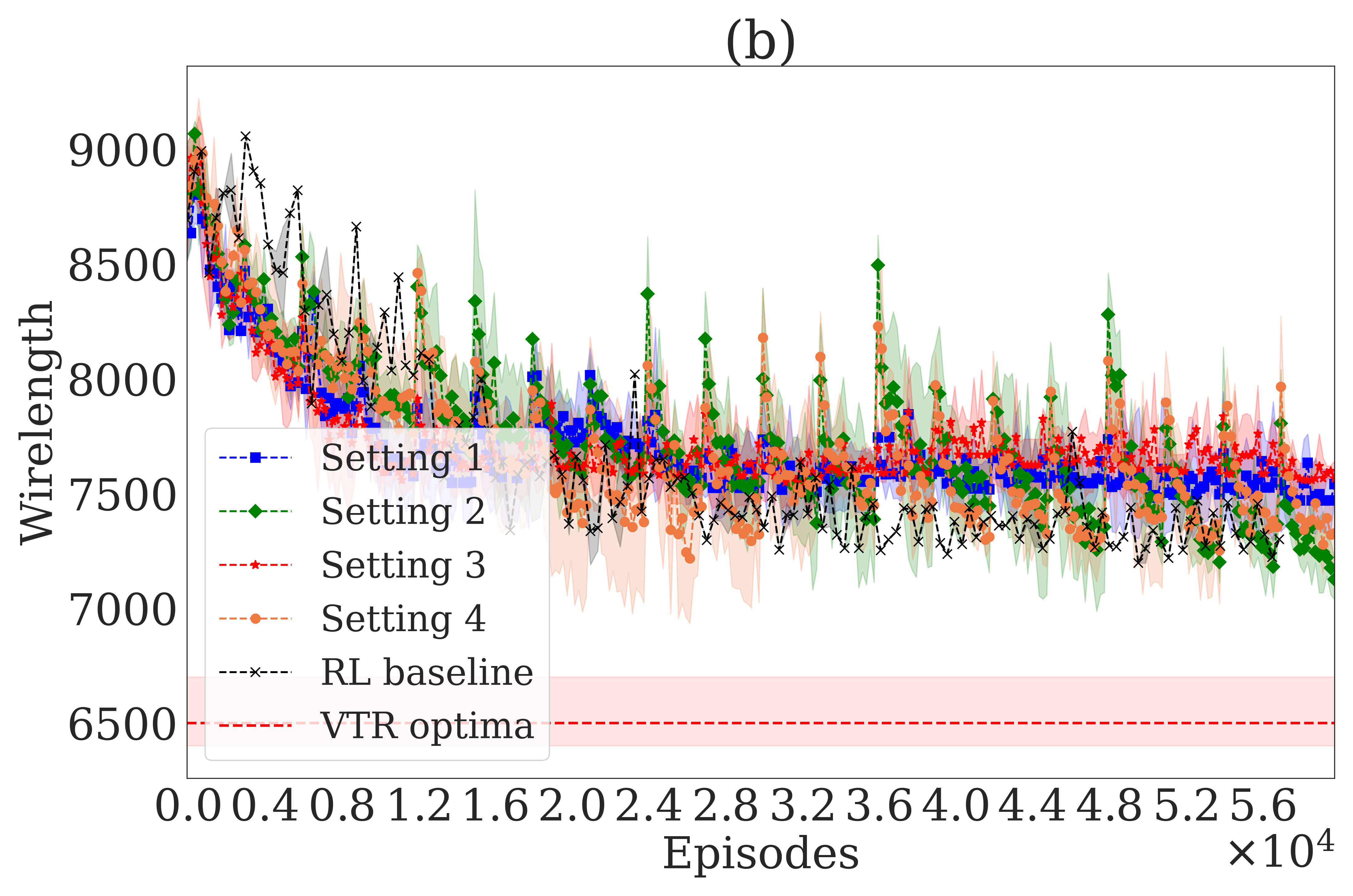}
    \includegraphics[width=0.65\columnwidth]{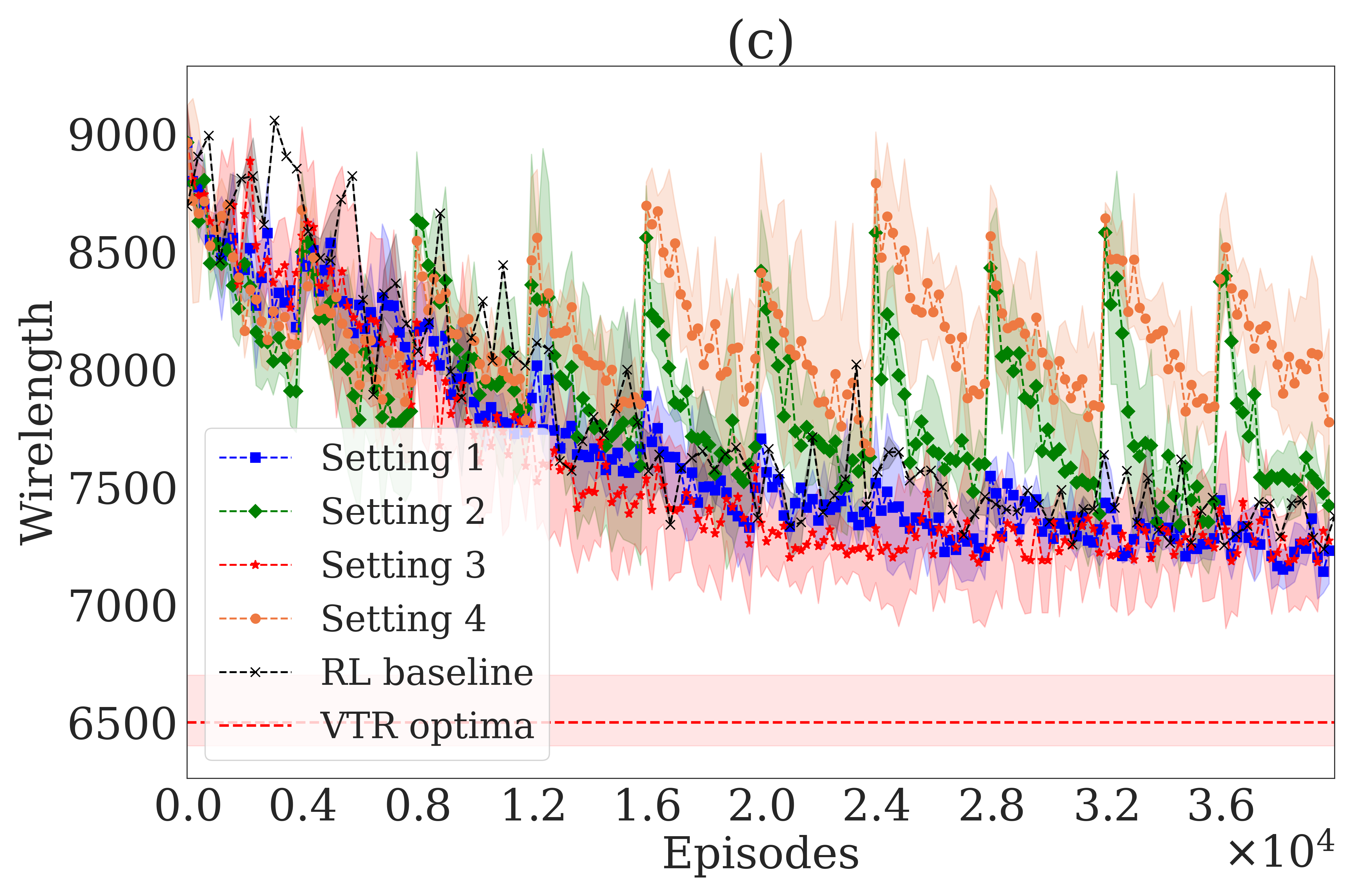}
\caption{The wirelength chart illustrates the policy performance curves during the training process. \textbf{(a)} the wirelength curves in 30-blocks decomposition; \textbf{(b)} the wirelength curves in 56-blocks decomposition with 4 granularity; \textbf{(c)} the wirelength curves in 30-blocks decomposition with 2 granularity.}
\label{decomposition_wirelength}
\end{figure*}

In our decomposition experiments, we set our RL methodology without decomposition as the baseline. 

\subsection{30 Blocks Decomposition} 
In 30 block decomposition experiments, each subtask (placing 15 blocks) undergoes training for 3000 episodes, with the training process repeated across 5 iterations. In Fig.~\ref{decomposition_wirelength} (a), when compared to our RL baseline, decomposition experiments exhibit a temporary spike in wirelength each time a subtask changes, particularly evident in settings 2 and 4. The wirelength peaks occur as the agent requires a policy transfer to adapt to the new state transitions and state spaces following each switch.
However, when comparing settings 2 and 4 with settings 1 and 3, we see that settings 1 and 3 tend to have worse performance. Moreover, the wirelength peaks with each subtask change tend to level off as training progresses. Fig.~\ref{decomposition_entropy} (a) reveals that the entropy of settings 1 and 3 progressively reduce over time and ultimately converges towards 0, which suggests that fully reusing model weights (policy) can accelerate the convergence speed but does not ensure convergence to global optima. The reused policy limits the agent to explore the environment, leading to a negative policy transfer effect that potentially traps the agent in inferior local optima. 


\begin{table}[tt]
\begin{minipage}{\columnwidth}
    \begin{center}
    \caption{The fifth row and the last row are our RL baseline for 30 blocks and 56 blocks, respectively. Granularity refers to the number of subtasks. Avg wirelength is the mean and standard deviations calculated from 3 different seeds. Best is the top performance among the three seeds. The bold rows indicate optimal settings for 30-block, 56-block with 2 granularity, and 56-block with 4 granularity decomposition, respectively.}
    \begin{tabular}{p{0.8cm}p{1cm}p{0.8cm}p{0.7cm}p{1cm}p{1.2cm}p{0.5cm}}
\hline
\hline
\scriptsize\textbf{\# blocks} & \scriptsize\textbf{\# policy} & \scriptsize\textbf{$W_{decis}$ reuse} &  \scriptsize\textbf{setting} & \scriptsize\textbf{granularity} & \scriptsize\textbf{avg wirelength} & \scriptsize\textbf{best}\\
\hline
\hline
\multirow{5}{0.8cm}{30} & 2 & T & 1 &2 & 6974$\pm$87 & 6749\\

 & \normalsize\textbf{2} & \normalsize\textbf{F} &  \normalsize\textbf{2} &  \normalsize\textbf{2} &   \normalsize\textbf{6795$\pm$160} &  \normalsize\textbf{6546}\\

 & 1 & T &  3& 2  & 7222$\pm$126& 7108\\

 & 1 & F & 4 &2 &  6852$\pm$27& 6638\\

 & 1 & NA & NA & NA & 6988$\pm$98& 6884\\
\hline
\multirow{9}{0.8cm}{56} &  \normalsize\textbf{2} &  \normalsize\textbf{T} &  \normalsize\textbf{1} &  \normalsize\textbf{2} &  \normalsize\textbf{7192$\pm$131} &  \normalsize\textbf{7081}\\

 & 2 & F & 2 & 2 & 7308$\pm$111& 7013\\

 & 1 & T & 3 &2 & 7226$\pm$234& 6971\\

& 1 & F & 4 &2 & 7721$\pm$185& 7193\\
& 4 & T & 1 &4 & 7499$\pm$162& 7334\\

 & \normalsize\textbf{4} & \normalsize\textbf{F} & \normalsize\textbf{2} &\normalsize\textbf{4} & \normalsize\textbf{7169$\pm$90}& \normalsize\textbf{7075}\\

& 1 & T & 3 &4 & 7561$\pm$85& 7427\\

 &  1 &  F &  4 & 4 &  7299$\pm$105&  6950\\

& 1 & NA & NA & NA & 7591$\pm$151& 7265\\
\hline
\end{tabular}
\label{tab2}
\end{center}
  \end{minipage}
\end{table}

\subsection{56 Blocks Decomposition} 
\begin{figure*}[h]
\centering
    \includegraphics[width=0.65\columnwidth]{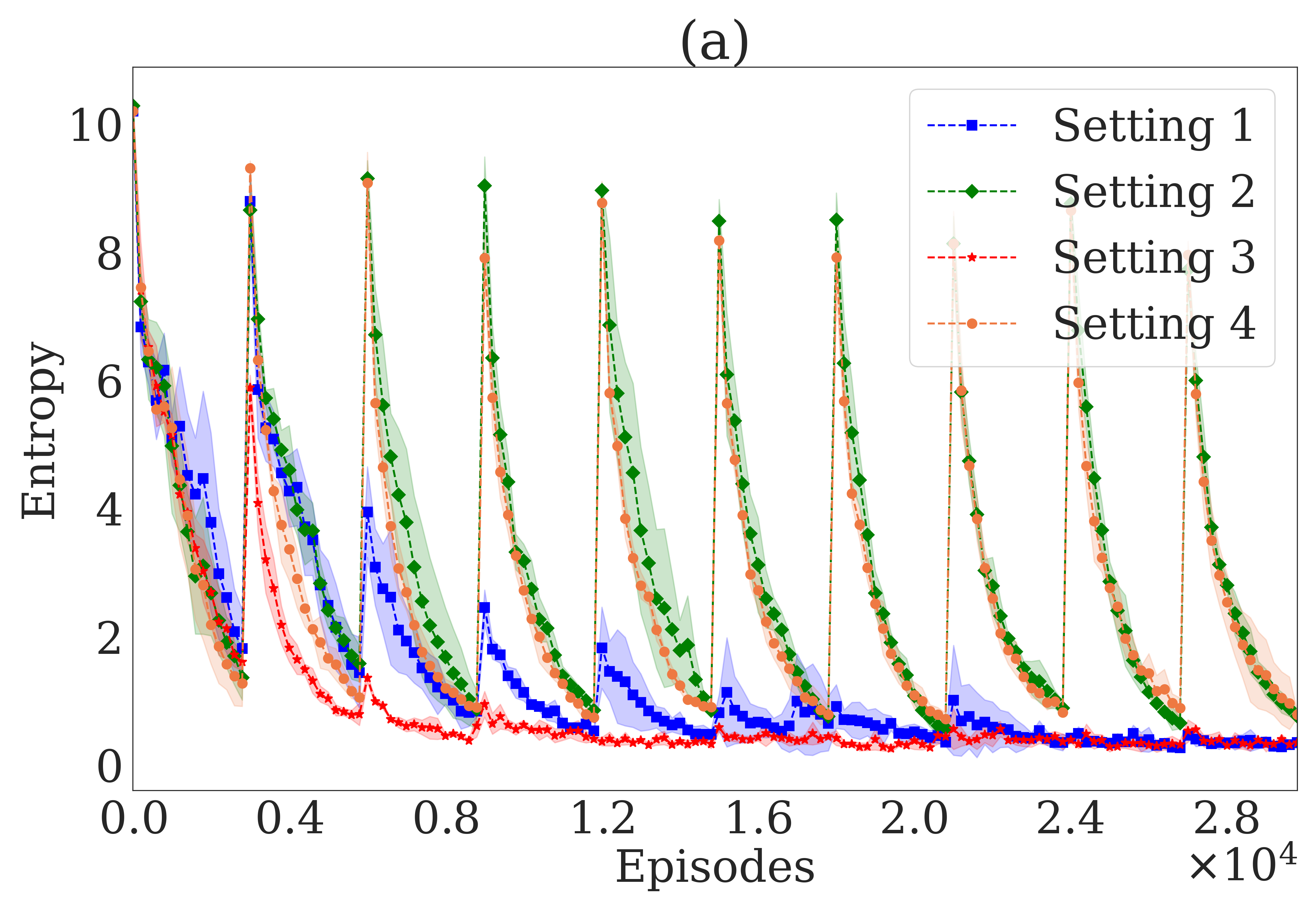}
     \includegraphics[width=0.65\columnwidth]{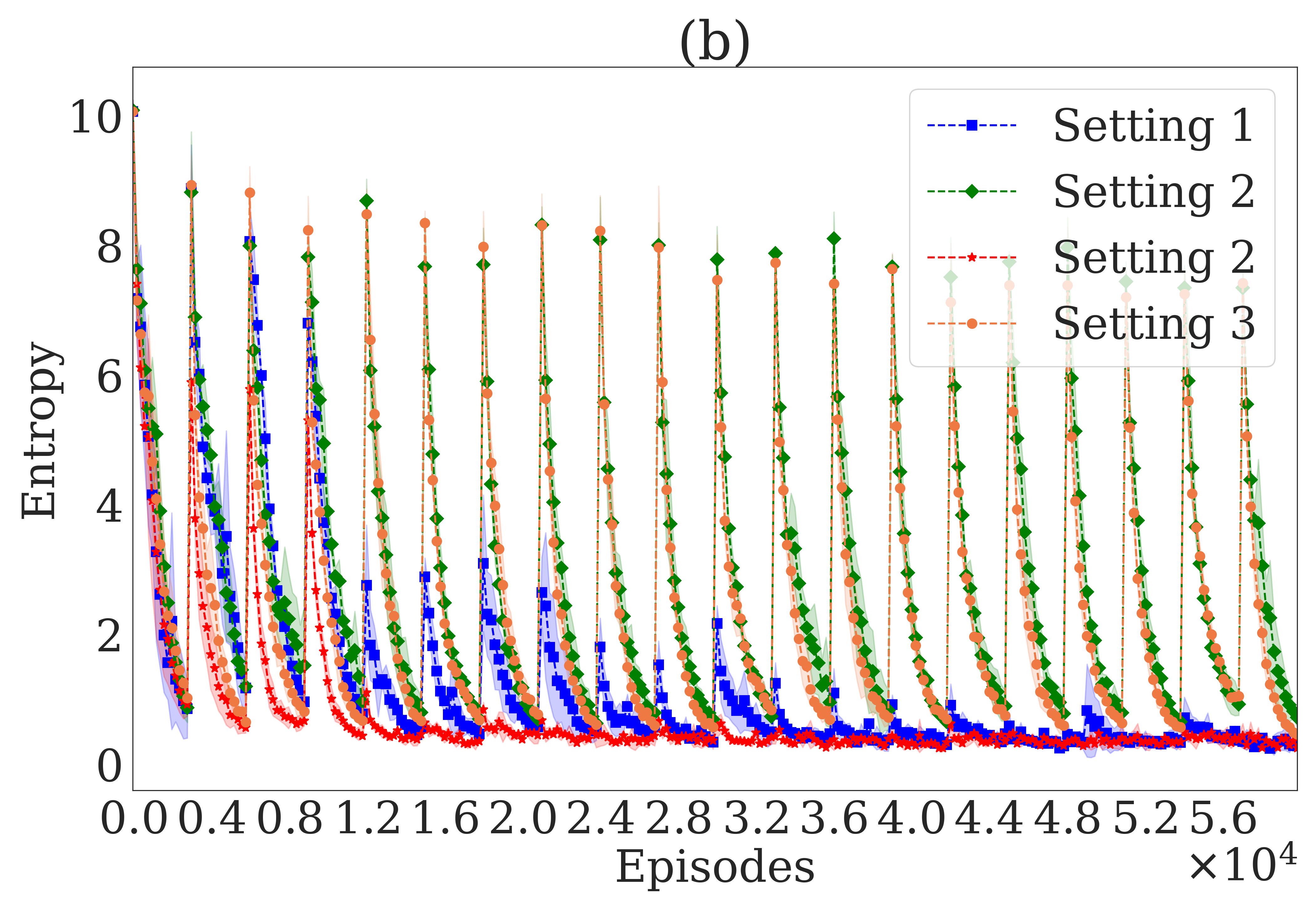}
      \includegraphics[width=0.65\columnwidth]{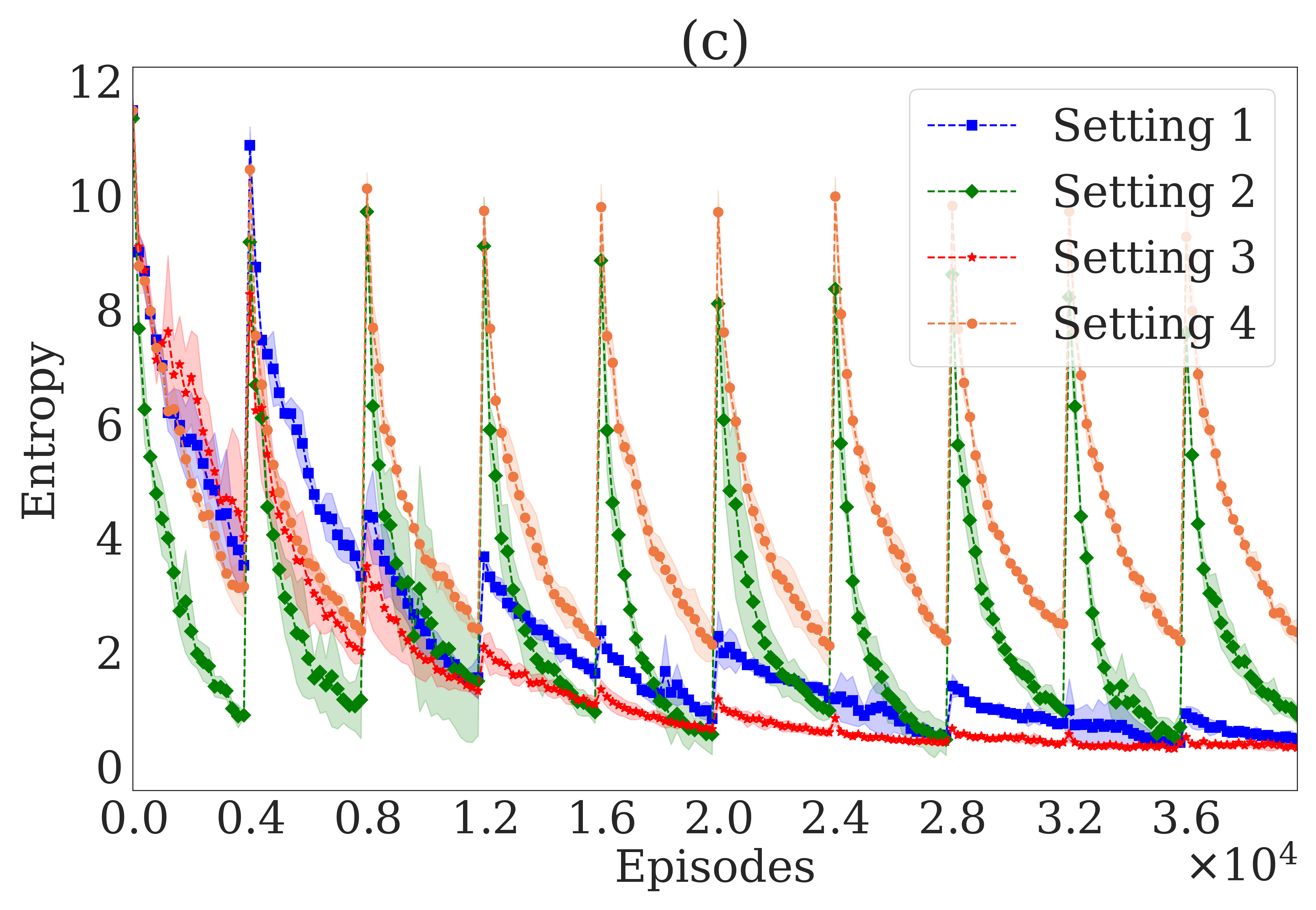}
\caption{The entropy chart illustrates the policy entropy curves during the training process, highlighting how the policy exploration evolves with 30 blocks (a), 56 blocks with graularity 4 (b), and 56 blocks with granularity 2 (c).}
\label{decomposition_entropy}
\end{figure*}

In a 56 block decomposition, each subtask placing 14 blocks trains for 3000 episodes, with the training process repeated across 5 iterations. From Fig.~\ref{decomposition_wirelength} (b) and Fig.~\ref{decomposition_entropy} (b), the performance is very similar to that of 30 blocks. In both 30 and 56 block decompositions, reusing all model weights results in negative policy transfer. 


\subsection{Granularity}
In our 56 block decomposition experiments, we further explore the impact of granularity on the performance by conducting extra experiments with each subtask placing 28 blocks, where each undergoes training for 4000 episodes, repeated over 5 iterations. Fig.~\ref{decomposition_wirelength} (c) shows that reusing all network weights performs better than reusing only the representation layer weights. This finding contrasts with the results of decmposing 56 blocks into 4 subtasks. 
We hypothesize this is due to the differences in the task difficulties and the learnability of the model. As shown in Table~\ref{tab1}, the exploration space of 14 block subtasks is smaller than 28 block subtasks, leading to a higher risk of policy overfitting after reusing all weights. This overfitting is the underlying reason for the negative policy transfer observed in 14 block subtasks, which hinders the agent from exploring the environment effectively.

Our experiments reveal that both the granularity of subtasks and the choice to reuse decision layer weights play crucial roles in performance. Specifically, in subtasks with finer granularity, not reusing the decision-layer weights promotes agent exploration. In contrast, subtasks with larger graularity show that reusing decision layer weights can improve performance. In general, results in Table~\ref{tab2} 
demonstrate that the decomposition training paradigm can improve performance compared to the baseline. 



\section{Discussion and Future Work}
In our decomposition setting, we explored two approaches: reusing the decision layer weights and not reusing them. While reusing the decision layer weights introduces more prior knowledge, it may also increase the risk of overfitting and limited exploration. On the contrary, not reusing the decision layer weights fosters exploration, but disregards valuable learned policy knowledge. Identifying a method to partially reuse decision layer weights to leverage prior policy knowledge while still encouraging agent exploration may improve performance.

The large search space can lead agents to become trapped in local optima. Efficient exploration is critical to overcome these challenges. Others \cite{lai2022maskplace} leverage reward engineering by using incremental estimated wirelength as step rewards, transforming sparse rewards environments to dense ones. This approach provides agents with frequent feedback, facilitating more effective policy evaluation and updates. Random Network Distillation (RND) \cite{cheng2021joint} can also generate intrinsic rewards. This strategy aims to enhance exploration by rewarding the agent for discovering new aspects of the environment. We decompose full placement problems into manageable subtasks, enabling agents to navigate smaller search spaces more effectively. These approaches aim to boost the agent's learning efficiency within the expansive search space. In the future, we also plan to combine a divide-and-conquer strategy with better RL exploration algorithms to facilitate more efficient training for the agent.

A good placement should not only optimize wirelength but also account for critical time delays, congestion, and processing time.
Recognizing that the definition of an ``optimal" placement varies based on different objectives, it often necessitates considering tradeoffs among these competing metrics. One could approach this by linearly combining different metrics or formulating the problem as a multi-objective RL task to optimize multiple conflicting objectives simultaneously. In light of this, a significant future direction for our research will expand our optimization objectives to include time delay and processing time for FPGA placement. Moreover, we aim to boost the versatility of our method by conducting experiments with a wide array of netlists, which will allow us to evaluate its performance across different scenarios. This step is intended to provide a more thorough understanding of how our approach performs in various settings, improving our understanding of its overall effectiveness.

\section{Conclusion}
This paper introduces an RL model for FPGA placement problems and a divide-and-conquer approach to optimize chip placements. This paper formulated the FPGA placement problem as an MDP and used an RL algorithm to learn a placement policy. Much of our efforts focused on constructing an appropriate state representation as the state used for prior ASIC placement methods was not applicable. To address the large space and sparse rewards, we proposeed a decomposition training paradigm. Although our approach may not yet outperform VTR, we have achieved promising preliminary results. Our work represents a significant step forward in leveraging RL for FPGA placement, and we hope others will contribute to further advancements in this exciting area. 

\section*{Acknowledgments}

This work has taken place in the Intelligent Robot Learning (IRL) Lab at the University of Alberta, which is supported in part by research grants from the Alberta Machine Intelligence Institute (Amii); a Canada CIFAR AI Chair, Amii; Digital Research Alliance of Canada; Huawei; Mitacs; and NSERC. 
The authors thank Qianxi Li for input on earlier versions of this work, as well as Peter Chun and Mark Bourgeault for their guidance throughout this research project.

\bibliographystyle{IEEEtran}
\bibliography{ref}

\end{document}